\documentclass[10pt,a4paper,oneside,onecolumn]{article}
\usepackage{graphicx}
\usepackage{amssymb}
\begin{document}
%%%%%%%%%%%%%%%%%%%%%%%%%%%%%%%%%%%%%%%%%%%%%%%%%%

%%%%% AUTHORS - PLACE YOUR OWN COMMANDS HERE %%%%%

% Please keep new commands to a minimum, and use \newcommand not \def to avoid
% overwriting existing commands. Example:
%\newcommand{\pcm}{\,cm$^{-2}$}	% per cm-squared

%%%%%%%%%%%%%%%%%%%%%%%%%%%%%%%%%%%%%%%%%%%%%%%%%%

%%%%%%%%%%%%%%%%%%% TITLE PAGE %%%%%%%%%%%%%%%%%%%

% Title of the paper, and the short title which is used in the headers.
% Keep the title short and informative.
\section*{ Galactic %coincidences in Earth's %0.2 - 2 mHz 
hum}

% The list of authors, and the short list which is used in the headers.
% If you need two or more lines of authors, add an extra line using \newauthor

{\bf Francesco Mulargia \\
}
% List of institutions
{Dipartimento di Fisica e Astronomia, 
Universit\`a  di Bologna, viale Berti Pichat 8
40127 Bologna, Italy (francesco.mulargia@unibo.it)\\
}

% Abstract of the paper
\begin{abstract}

A number of earth's tremor spectral peaks  
show a persistent narrow bandwidth 
incompatible with any geophysical or instrumental origin. 
These peaks, located at frequencies lower than a few mHz, are in principle consistent  
with the earth strain waves induced by monochromatic gravitational waves. 
Exploring this 
hypothesis under the current cosmological constraints yields that the tremor peaks below 2 mHz 
are in apparently significant coincidence with the theoretical 
 emission of two binary systems each consisting  of a small main sequence star with mass 
 $\sim 10^{-1} M_{\odot}$, captured by Sgr A* in a close orbit.

\end{abstract}

%%% Keywords should appear after the \end{abstract} command. 
%%% See the online documentation for the full list of available subject
%%% keywords and the rules for their use.
%\keywords{gravitational wave detectors and experiments; noise; seismicity; earth surface waves and free oscillations}binary system
%
{\it Keywords:}
gravitational wave detectors and experiments; noise; seismicity; surface waves and free oscillations;
%\pacs{  
04.80.Nn, 05.40.Ca, 91.30.Dk, 91.30.Fn

\section*{Introduction} %\label{sec:intro}

Seismic tremor, i.e., the continuous background motion of the earth apparent in seismic and gravimetric records, also known with the name of \em  earth's  hum\em, shows  a
number of
narrow "anomalous" spectral peaks significant 
beyond 3$\sigma$
(Nawa et al. 1998; Tanimoto et al., 1998; Thomson \& Vernon 2015). These peaks, which are located in the sub-mHz and mHz band -- and occur in  correspondence with solar acoustic modes (see Table \ref{eigen}) -- can hardly %not 
be ascribed to instrumental effects, nor can they be 
 produced by the earth elastic response to the continuous broadband 
excitation by infragravity sea waves, which constitute the main tremor source 
(Webb 2008). 

Explanations have been advanced in terms of some yet unknown acoustic-magneto-elastic effect of the sun on the earth (Thomson  \& Vernon 2015), and of an elastic excitation of the earth from persistent monochromatic gravitational waves of cosmic origin, which would equally affect the earth and the sun, given their comparable gravitational cross-sections (Mulargia 2017).
 Here, we further explore the latter 
 hypothesis taking into account the known cosmological evidence and focusing our attention at frequencies below 2 mHz, where most of the unexplained peaks occur.

\section*{The earth as a low frequency  gravitational  detector}
    
The viscoelastic gravitational cross section of the earth $\Xi$  
can be written as a function of its 
mass $M$ and geometric area $A$   
as (ibid.) 

\begin{equation}
\Xi = \frac{4 \pi^2}{15} \big( \frac{r_g f}{c} \big) Q A
\label{cross}
\end{equation}

\noindent where $r_g$ is the Schwarzschild radius,
$f$ the wave frequency and $G$ the gravitational constant, 
 with $\Xi$ attaining the quite large value of several $ 10^3$ m$^2$ per mHz. 
Combined with this, two intrinsic amplification mechanisms nominate the earth as a %potential 
 low-frequency gravitational wave detector. 

 First, its \em elastic response \em to gravitational wave excitation at frequencies proximal to %in correspondence of 
 its eigenmodes produces an amplification by a factor $q$. %of the order of $10^2$. 
 In fact, the earth behaves as a damped harmonic oscillator with quality factor  $Q \sim 200 \doteq 400$ (see table 1) which, once the system has attained dissipative equilibrium, resonates amplifying the excitation signal. Note that equilibrium is attained after $Q/2 \pi$  cycles, i.e., in the mHz frequency band we consider, after no less than several hours, restricting  detectability to monochromatic and persistent signals.

Second, the main source of terrestrial tremor is constituted by oceanic wave-wave and wave-bottom interactions  (e.g., Webb, 2008), which
are 1) highly nonlinear, 2) acting upon an excitable system dominated by noise and 3) ruled by thresholds (e.g., Schwartz \& Fenton 1982). Such systems are known to produce \em stochastic amplification \em under rather general conditions (Benzi et al 1981; McNamara et al. 1988; Jung \& H\"anggi 1991). 
Stochastic amplification is the -- somehow counterintuitive -- statistical facilitation of the transition 
to a higher energy state 
by the addition of random noise,
and is known to affect a wide variety of nonlinear noisy phenomena ruled by thresholds, from neuron firing to climate cycles, digital image dithering,  superconducting quantum interference, bistable lasers,  etc.  
 (for a quite comprehensive list see e.g., Gammaitoni et al. 1998).  The 
relevance of stochastic amplification to the present case
stems from the experimentally observed modulation of the narrow spectral peaks by the background noise level  (Nawa et al. 1998; Tanimoto, 1998; Webb, 2008).

Hence, the measured  
tremor monochromatic displacement $x$ at the frequency $f_0$ is equal to
the gravitational excitation  
displacement $u$, amplified by an elastic response factor $q$  (tied to the proximity of $f_0$ to an eigenmode and its related $Q$) and by a stochastic factor $K$ (tied to the background noise level and dynamics), i.e., $x =  u   q   K $.

Due to the perturbation induced by earth daily rotation, the gravitational wave sources are likely to be fewer than the number of tremor peaks. In fact, the earth rotates at the frequency $F=0.01157$ mHz and perturbs any 
excitation
at frequency $f$ by inducing a Zeeman-like splitting of $f$ into the singlets  $ f, f \pm  F ,  f \pm 2 F$. 
Such a splitting, called \em Coriolis splitting\em,
 is routinely observed right after large earthquakes, which make the whole earth "ring" (Dahlen \& Tromp 1998), and is also apparent in narrow tremor spectral peaks (Thomson \& Vernon 2015; Mulargia 2017). 
 
\section*{Compatible sources}

Since the gravitational wave (from now on GW) emission power depends on the time variation of the source quadrupole moment, a most effective GW source
 consists of massive compact rotating binary systems, with orbital eccentricity $e$ also playing a substantial role 
(Peters \& Mathews  1963; Pierro et al. 2001).

From the above arguments and Kepler's 3rd law we may define the set of binary systems of total mass $M_b$ illuminating the earth with a monochromatic GW power flux $\Psi$ at the frequency $f$ from a distance $d$ as  (Mulargia, 2017) 
\begin{equation}
\frac {({f M_b})^{10/3}} {d^2} \simeq  \Psi \frac {c^5} {(2 \pi G )^{7/3}}
\label{Phi}
\end{equation}
\noindent which provides the total mass  - distance source ratios compatible with a measured power flux at a given frequency. 

\section*{Experimental vs theoretical spectra } 

We consider here (Table 1) the 
unexplained tremor narrow spectral
peaks located below 2 mHz, for which consistent evidence exists relative to time intervals up to three continuous years (Nawa 1998; Thomson \& Vernon 2015). 

  Among the best known gravitational source candidates 
are the 
binary systems consisting of a compact object captured in close orbit by a supermassive black hole (from now on SMBH), the 
nearest of which is obviously Sgr A*. 
Extensive numerical simulations of the capture and orbital evolution of such systems (Sigurdsson \& Rees 1997; Freitag 2003; Barack \& Cutler 2004) show that most GW emission from captured flying-by compact objects --  be them stellar black holes (from now on BH), neutron stars (from now on NS), white dwarfs (from now on WD) and small main sequence stars with mass $M\sim 10^{-1} M_{\odot} $ (from now on SMSS) -- occur in the 0.1 - 5 mHz frequency band, and with a comparable power in several %the first 5 
spectral components.     

Namely, each compact body
flying by the SMBH at a distance $ 5 - 20 r_g$ 
is captured in a highly elliptical orbit,
with initial 
$(1-e) \lesssim 10^{-4}$ (e.g., Freitag, 2003, Fig.1).
In such an orbital capture, the moment of inertia of the SMBH-compact body binary %system 
 experiences a strong change, emitting  gravitational waves at the expense of the system 
energy.  
Initially, the GW emission occurs in
"bursts" near perihelion,  while 
orbits become progressively less eccentric, at a rate 
$de/dt \sim  {M_{\rm c.body}} /{M_{\rm SMBH}^2}$. 
However, most gravitational energy is not emitted at plunging, but throughout the whole orbital evolution, and consists of monochromatic GW wavetrains 
in the $10^{-5} \doteq 10^{-2} $ Hz frequency band, with an emission life  
 longer the lighter the compact body. In particular, a SMSS-SMBH binary would orbit 
 with eccentricities around 0.5  for most of its life, emitting stably with little frequency drift and with power increase factors of 3-7 with respect to a circular orbit
Freitag 2003; Barack \& Cutler 2004). 

  The inspiral orbiting would end for all such binaries with a plunging of the compact body on the SMBH, accompanied by a high frequency GW chirp (e.g., Abbott et al. 2016).  However, this would  
not occur for the SMSS--SMBH binaries, which die by slow tidal disruption of the SMSS during several decades (Rees 1988; Lin et al. 2017).

 \section*{Discussion }

\begin{figure}
%\centering
\includegraphics[width=\textwidth]{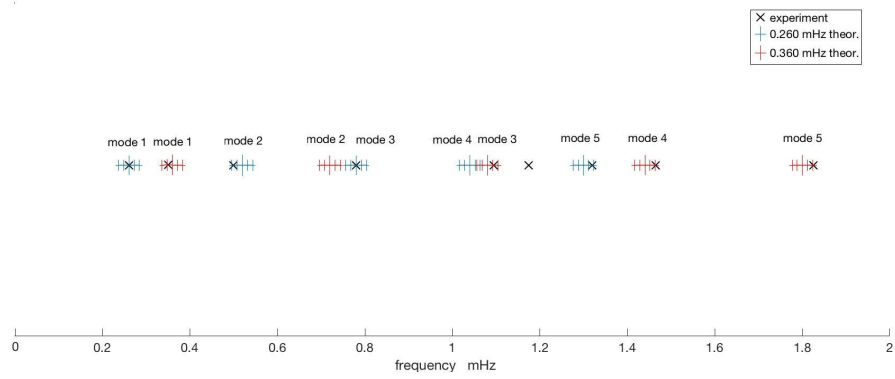}
\caption{ The earth's narrow tremor spectral peaks of table 1 are indicated by black x's. The first five theoretical modes of monochromatic emissions at 0.260 and 0.360 mHz are reported by blue and red crosses, together with the 
Coriolis %$\pm F, \pm 2 F$ 
 frequency bands (see text). }
\label{multiples}
\end{figure}

Let us compare the experimental unexplained narrow tremor peaks of table 1  
with the theoretical emission of the mentioned GW sources. 
According to the above amplification arguments, we expect the 
lowest frequency signal by the earth fundamental eigenmode, which is $_0S_{2}$ at 0.309 mHz.
In fact, the lowest fequency peaks are in its proximity, at $f_0=0.260$ and $f_0=0.360$ mHz. We simply hypothesize that they are produced by two GW sources, and the other peaks are their spectral overtones up to order 5 (see Fig. \ref{multiples}). We consider the Coriolis frequency splitting induced by Earth rotation at  $f_0,  f_0 \pm  F ,  f_0 \pm 2 F$, with $F=0.01157$ mHz and the additional frequency perturbations induced by the lateral heterogeneities of the earth, i.e., its ellipsoidal shape and the inhomogeneities in its internal structure (Dahlen \& Tromp 1998).  Since these effects are of the order of a few $\mu$Hz, we account for them altogether by considering $f_0 \pm 2 F$ bands around each theoretical mode. In doing so, we find a theoretical - experimental match for 8 tremor spectral peaks out of 9: 
 two of the theoretically predicted peaks are absent, respectively the 4th (at 1.040 mHz) of the $f_0=0.260$ mHz, and the 2nd of the $f_0=0.360$ (at 0.720 mHz), while the experimental peak at 1.175 mHz is left theoretically unexplained.  

Assuming the experimental peaks to be independent and identically distributed according to a uniform distribution with mean equal to the empirical mean, such a match can be ascribed to random chance with a probability 
$P \simeq  {{m \choose r} {int(T/\Delta t) - m \choose {n-r}}} / {{int(T/\Delta t) \choose n }}  $. Here 
$r$ is the number of matches, $int$ the integer part and $\Delta t$ the frequency bin
$\Delta t = 4 F$, with $m$ the number of experimental peaks, $n$ the number of theoretical modes, and
 $T$ the total frequency interval considered. Given that $m=9, n=10, r=8, T=2, F=0.01157$, we have $P\sim 10^{-6} $,  corresponding to a random chance level $\sim 5 \sigma$.  Incidentally, we note also that the 2.614 mHz tremor peak considered by Mulargia (2017) would occur by the 10th spectral overtone of the $f_0=0.260$ mHz fundamental.
      
The 0.260 mHz  peak is about $19$\% off the closest earth eigenmode, $_0S_2$, which occurs at 0.309 mHz with a $Q=463$,  so that only a modest elastic response amplification can be expected (a factor $\sim 3$). 
The other spectral peaks 
share a closer proximity to an earth eigenmode, 
mostly occurring within a few percent  (table \ref{eigen}).  Given that these  eigenmodes have $Q \gtrsim 200$ values,
an amplification by elastic response of the incoming GEW by a factor $q \gtrsim 10 $ can  
be expected (Mulargia 2017).  A similar factor $K \gtrsim 10 $ can be expected for stochastic amplification in light of the narrow spectral peaks modulation by background tremor, which raises them to 
 amplitudes at least one order of magnitude larger in winter for stations in the northern hemisphere, when oceanic noise greatly increases. Hence, 
 GEW amplitude $u$ can be estimated from measured tremor amplitude as $ u = x/(qK) \lesssim 
 %10^{-3} %\doteq 
 10^{-2} x$.  
 
Given the measured displacement power spectral density $P(x; f_0)$,
the displacement amplitudes $x$ at $f_0$ can be written 
as (cf.  Mulargia \& Kamenshchik 2016) $x =  \sqrt {2 P(x) S_{BW}}$,  where 
  $S_{BW} = f_u - f_l$ is the spectral bandwidth,  and $f_u$, $f_l$ respectively the upper and lower corner frequencies around $f_0$. Considering, for example,  the peak at 0.360 mHz, we
  have $P(x) \sim 10^{-10}$ m$^2$/Hz in "quiet" seismic records (Peterson 1993; Castellaro \& Mulargia 2012) and  $S_{BW} \sim 10^{-6}$ Hz, $x \sim 10^{-8}$ m, while gravimetric records, with the superior sensitivity of superconducting gravimeters below 1 mHz, yield measured amplitudes $x \sim 10^{-9}$ m (Widmer-Schnidrig 2003; Rosat et al. 2003). These stand for
  GEW amplitudes $ u \sim 10^{-11} %\doteq 10^{-10} 
  $ m, or, given the average velocity of elastic $P$ waves in the earth mantle of $\sim 10^4 $ m/s,  rms strains
  $h_{rms} \sim 10^{-19}$, 
coincident in both frequency and amplitude with those theoretically calculated for the gravitational emission of 
 compact body - Sgr A* binary systems (Freitag 2003, Fig. 2;  Barack \& Cutler 2004, Fig. 11).  

 Binary systems with compact bodies other than SMSS would rapidly evolve into merging, emitting near monochromatic GW  for just a few decades, while the SMSS - Sgr A* systems would emit with very small frequency drift for $ \gtrsim10^4$ years (ibid.). In addition, the current cosmological, microlensing and %black hole 
 density constraints (Alcock et al. 1996; Ado et al. 2014, Abbott et al. 2016),  suggest an expected number of GW monochromatic signals with $h_{rms} \sim 10^{-19}$, i.e.,  detectable by one year long eLISA
 future missions, in number $\ll 1$ for binaries formed by Sgr A$^*$ with a BH and a NS, $\sim 1$ with a WD and $\gtrsim 10$ with a SMSS of mass $M\sim M_{\odot}/20 $ (Freitag 2003). 
 Since such a 
resolving power in the mHz band is likely to be achieved by adequately long  
records of the earth's tremor (Mulargia \& Kamenschick 2016; Mulargia 2017), like the ones used to identify the anomalous spectral peaks in table 1,
the above coincidences appear consistent also with the expected source numerosity.

 \section*{Conclusions}

Gravitational waves illuminating the Earth may induce detectable seismic tremor. Thanks to its size and structure, the gravitational wave "stopping power" of the earth is  considerable, with a cross section $\sim 10^4$ m$^2$ in the mHz band. This, together with its elastic response amplification, the stochastic amplification provided by ocean dynamics, and the intrinsic stability of seismographs and gravimeters, makes the earth a low-frequency gravitational wave detector with a  $h_{rms} \sim 10^{-19}$ strain sensitivity in the mHz band. 

A number of earth's tremor narrow spectral peaks, measured 
by independent observers and not amenable to a terrestrial origin, 
show remarkable coincidences with those theoretically calculated for the gravitational wave emission of compact bodies orbiting Sgr A*, following their capture. 
While these compact bodies could be either a stellar black hole, a neutron star, a white dwarf, or a small main sequence star, their 
relative cosmological abundance 
suggests the latter as much more likely. 
In particular, the earth tremor "anomalous" spectral peaks below 2 mHz 
are in apparently significant 
coincidence with the gravitational wave emission of two binary systems consisting of a star with mass of the order of $10^{-1} M_{\odot} $ captured by Sgr A* in a close orbit.
Such binaries would not die by merging, 
and therefore would not lead to any signal detectable by terrestrial gravitational interferometers, which operate at frequencies above 10 Hz.

\section*{Acknowledgments}
The author is indebted to Alberto Sesana, Luca Ciotti, Giancarlo Setti,  Bruno Marano and Philip B. Stark for  enlightening discussions.

\newpage
%\begin{thebibliography}{}
\section*{REFERENCES}

{ $÷÷\quad$  } Abbott, B., et al. (LIGO Sci. %entific 
 Coll. %aboration and 
 \& Virgo Coll.)%aboration)
 , 2016, 
 %Observation of gravitational waves from a binary black hole merger, 
Phys. Rev. Lett., %}, {\bf
%PhRvL, 
116,  %(2016) 
061102%.

\smallskip%{[1]} 
Ado et al. (Planck coll.),  2014,  %results. XVI. Cosmological parameters
Astron. Astrophys., 571,  A16

\smallskip%{[2]}
Alcock, C., Allsman, R. A., Axelrod, T. S., et al. 1996, 
% The macho project: microlensing detection efficiency, 
461, 84

\smallskip%{[3]}
Benzi, R., Sutera, A. , \& Vulpiani, A., 1981, J. Phys. A 14, L453

 \smallskip%{[4]}
Barack, L., \& Cutler, C., 2004, %LISA capture sources: Approximate waveforms, signal-to-noise ratios, and parameter estimation accuracy?. 
Phys, Rev. D 69, 082005

\smallskip%{[5]} 
Castellaro, S., \& Mulargia, F., 2012, %  A Statistical Low Noise Model of the Earth,
 Seismol. Res. Lett., 83, 39%-48. 
 
\smallskip%{[6]}
Dahlen, F. A.  \&  Tromp, J. T. 1998, 
Theoretical global seismology,
(Princeton: Princeton Press)%1024 pp.

\smallskip%{[7]}
Dziewonski, A. M.  \&  Anderson,  D. L., 1981, 
%Preliminary reference Earth model, 
%{\it 
Phys. Earth Plan. Interiors, %}, 
25, 297%-356.

\smallskip%{[8]}
Freitag. M., 2003, 
% Gravitational waves from stars orbiting the Sagittarius A* black hole,
Astrophys. J., 583, L21%-L24.

\smallskip%{[9]}
Gammaitoni, L., H\"angii, P., Jung, P.  \& Marchesoni, F., 1998, 
%Stochastic resonance, 
%{\it 
Rev. Mod. Phys., %}, 
70,  223%-287.

\smallskip%{[10]}
Jung, P. \& H\"angii, P., 1991,
%Amplification of small signals via stochastic resonance
Phys. Rev. A, 44, 8032

\smallskip%{[11]}
Lin, D. et al., 2017,
%A likely decade-long sustained tidal disruption event
Nature Astron., Article 0033
% doi:10.1038/s41550-016-0033 

\smallskip%{[12]}
McNamara, B.,  \& Wiesenfeld, K., \& Roy,  1988,
% Theory of stochastic resonance
Phys. Rev. Lett., 60, 2626

\smallskip%{[13]} 
Mulargia, F., 2017, 
% Cosmic signatures in earth?s seismic tremor?
MNRAS 464, L11-L15 (2017)
doi:10.1093/mnrasl/slw180

\smallskip%{[14]}
Mulargia, F., \&  Kamenshchik, A., 2016, %Global seismic network as a GW Antenna, 
 Phys. Lett. A,  380, 1503%-1507.

\smallskip%{[15]}
Nawa, K., Suda, N., Fukao, Y.,  Sato, T.,  Aoyama, Y.,  \& Shibuya, K.,   1998, %. Incessant excitation of the Earth's free oscillations, 
Earth Plan. Space,
50, 3%-18. 

\smallskip%{[16]}
Peterson, J., 1993, % Observations and modeling of seismic backgroundnoise. 
USGS Open-File Report 93-322

\smallskip%{[17]}
Pierro, V., Pinto, I. M., Spallicci, A. D., Laserra, E., \& Recano, F., 2001,
MNRAS, 325, 358

\smallskip%{[18]}
Rees, M.J., 1988, % Tidal disruption of stars by black holes..
Nature, 333, 523

\smallskip%{[19]} 
Rosat, S., et al., 2003, 
Phys. Earth. Plan. Int., 140, 183

\smallskip%{[20]}
Schwartz, L. W., \&  Fenton, J. D., 1982, 
%Strongly nonlinear waves, 
Ann. Rev. Fluid Mech., 14, 39%-60.

\smallskip%{[21]}
Sigurdsson, S., \& Rees, M.J., 1997,
%Capture of stellar mass compact objects by massive black holes in galactic cusps
MNRAS, 284, 318%-326 (1997)

\smallskip%{[22]}
Tanimoto, T., Um, J., Nishida, K., \& Kobayashi, N, 1998,  
% Earth's continuous oscillations observed on seismically quiet days,
Geophys. Res. Lett., 25, 1553%-1556.

\smallskip%{[23]}
Thomson, D. J. \& Vernon, F. L. III, 2015, % Unexpected, high-Q, low-frequency peaks in seismic spectra,
%{\it 
 Geophys. J. Int., %}, 
202, 1690%-1710.

\smallskip%{[24]}
Webb, S. C., 2008, %
%The Earth's hum:the excitation of Earth normal modes by ocean waves,
%{\it 
Geophys. J. Int.,
%}, 
174, 542%-566.

\smallskip%{[25]}
Widmer-Schnidrig, R., 2003,
Bull. Seismol. Soc. Am, 93, 1370

\newpage
%\begin{longtable}{lcccccr}
\begin{table}
\centering
\caption{The frequency of the experimental earth tremor narrow peaks (Nawa et al. 1998; %Tanimoto et al. 1998; 
Thomson \& Vernon 2015) and neighbouring normal modes for the earth, spherical-S and toroidal-T, and the related $Q$ value (Dziewonski \& Anderson 1981) below 2 mHz. Neighbouring acoustic solar modes are also reported according to various sources (see suppl. to Thomson \& Vernon 2015). Note that their proximity to the tremor peaks is facilitated by their high density)
}
\label{eigen}
\bigskip
\begin{tabular}{lcccccr}

\hline
Tremor f, mHz  &  Earth Mode & f, mHz & Q &  Solar Mode & f, mHz\\
\hline
		   & & & & & \\

		   &$_0S_{2}$ & 0.309 & 463 & $ g_{2,-2} $ &  0.255 \\
		$\quad$0.260  & 	 &    &  & \\ %$\quad$0.250  & 	 &    &  &
		   &$_0T_{2}$ & 0.379 & 250 & $ g_{5,-6}$ & 0.257 \\

		   & & & & & \\

		    &$_0S_{2}$ & 0.309 & 463 & $ g_{5,-2}$ & 0.346 \\
		$\quad$0.350  & 	 &    &  &  \\ %$\quad$0.350  & 	 &    &  &
		   &$_0T_{2} $& 0.379 & 250 & $ f_2$ & 0.354 \\

		   & & & & & \\

		   &$_0S_{3} $& 0.469 & 421 & $ p_{2,2}$ &  0.513 \\
		$\quad$0.500  & 	 &    &  &  \\
		   &$_0T_{3}$ & 0.586 & 240 & $ p_{6,1} $& 0.493 \\
	
		   & & & & & \\

		   &$_0S_{5}$ & 0.840 & 356 & $ p_{10,2}$ &  0.778 \\
		$\quad$0.780  & 	 &    &  &  \\
		   &$_0T_{4} $& 0.766 & 228 & $ p_{22,1}$ & 0.784 \\
		   
		  & & & & & \\

		   &$_3S_{2}$ & 1.106 & 367 & $ P_{9,4}$ &  1.083 \\
		$\quad$1.095  & 	 &    &  &  \\ % $\quad$1.090  & 	 &    &  &
		   &$_0T_{6}$& 1.081 & 205 & $ P_{15,3}$ & 1.095 \\
		   
		   & & & & & \\

		   &$_1S_{4}$ & 1.173 & 271 & $ P_{19,3} $&  1.173 \\
		$\quad$1.175  & 	 &    &  &  \\ % $\quad$1.180  & 	 &    &  & 
		   &$_0T_{7}$ & 1.220 & 196 & $ P_{1,7}$ & 1.186 \\		   

  		  & & & & & \\
		   
		   &$_1S_{5}$ & 1.368 & 292 & $ P_{1,8} $&  1.330\\ 
		$\quad$1.320  & 	 &    &  &  \\
		   &$_1T_{2}$ & 1.321 & 256 & $ P_{11,5}$ & 1.321 \\		
		     		 
		 & & & & & \\
		   
		   &$_2S_{5}$ & 1.514 & 303 & $ P_{1,9} $&  1.473\\
		$\quad$1.465  & 	 &    &  &  \\ % $\quad$1.460  & 	 &    &  & 
		   &$_0T_{9}$ & 1.489 & 180 & $ P_{16,5}$ & 1.467 \\	

		 & & & & & \\
		   
		   &$_1S_{8}$ & 1.799 & 379 & $ P_{0,12} $&  1.823\\
		$\quad$1.825  & 	 &    &  &  \\ % $\quad$1.820  & 	 &    &  &
		   &$_0T_{12}$ & 1.859 & 163 & $ P_{5,10}$ & 1.826 \\	
		   
% 		& & & & & \\
%		   
%		   &$_4S_{4}$ & 2.280 & 290 & $ P_{0,12} $&  1.823\\in gravimetric and seismic data
%		$\quad$2.290  & 	 &    &  &  \\
%		   &$_1T_{8}$ & 2.281 & 232 & $ P_{5,10}$ & 1.826 \\			  
%			   
%		   		   
%		& & & & & \\
%		   
%		   &$_3S_{6}$ & 2.550 & 276 & $ P_{2,17} $&  2.620\\
%		$\quad$2.620  & 	 &    &  &  \\
%		   &$_1T_{10}$ & 2.620 & 223 & $ P_{11,14}$ & 2.620 \\
%
%
%		\ref{Phi}, & & & & & \\
%		   
%		   &$_3S_{8}$ & 2.822 & 264 & $ P_{1,19} $&  2.828\\
%		$\quad$2.830  & 	 &    &  &  \\
%		   &$_1T_{11}$ & 2.784 & 220 & $ P_{13,15}$ & 3.828 \\
%
%		   		   
%		& & & & & \\
%		   
%		   &$_3S_{11}$ & 3.221 & 250 & $ P_{1,22} $&  3.233\\
%		$\quad$3.220  & 	 &    &  &  \\
%		   &$_0T_{24}$ & 3.219 & 137 & $ P_{2,22}$ & 3.295 \\
%
%		& & & & & \\
%		   
%		   &$_1S_{17}$ & 3.494 & 159 & $ P_{3,23} $&  3.489\\
%		$\quad$3.490  & 	 &    &  &  \\
%		   &$_0T_{26}$ & 3.439 & 115 & $ P_{12,20}$ & 3.493 \\
%
%		& & & & & \\
%		   
%		   &$_3S_{13}$ & 3.508 & 241 & $ P_{6,22} $&  3.508\\
%		$\quad$3.510  & 	 &    &  &  \\
%		   &$_0T_{27}$ & 3.549 & 116 & $ P_{9,21}$ & 3.508 \\
%		   
		   
\hline
\end{tabular}
\end{table}
%\end{longtable}

\newpage

%\end{thebibliography}

\end{document}